\documentclass{iopart}
\usepackage{iopams}
\usepackage{graphicx}
\usepackage{color}
\begin{document}

\title[ J. Phys. D: Appl. Phys.]{Temperature Dependence of Spontaneous Switch-on and Switch-off of Laser Cavity Solitons  in Vertical-Cavity Surface-Emitting Lasers with Frequency-Selective Feedback}

\author{J. Jimenez, G.-L. Oppo and T. Ackemann}

\address{SUPA and Department of Physics, University of Strathclyde, Glasgow
G4 0NG, Scotland, UK}

\ead{jesus.jimenez-garcia@strath.ac.uk} \begin{abstract} A systematic experimental and numerical investigation of the conditions for the spontaneous formation of laser cavity solitons in broad-area vertical-cavity surface-emitting lasers with frequency-selective feedback by a volume Bragg grating is reported. It is shown that the switching thresholds are controlled by a combination of frequency shifts induced by ambient temperature and Joule heating. The gain level has only a minor influence on the threshold but controls mainly the power of the solitons. At large initial detuning and high threshold gain, the first observed structure can be a high order soliton. In real devices spatial disorder in the cavity length causes a pinning of solitons and a dispersion of thresholds. The experimental observations are in good agreement with numerical simulations taking into account disorder and the coupling of gain and cavity resonance due to Joule heating. In particular, we demonstrate that the existence of the traps explain the spontaneous switch on of the solitons,but do not modify the soliton shape significantly, i.e.\ the  observed solitons are a good approximation of the ones expected in a homogenous system.

\end{abstract}

\submitto{\JPD}
\maketitle

\section{Introduction}

Broad-area semiconductor lasers are attractive for high power applications but are limited by numerous instabilities leading to emission on many spatial modes and thus reducing the beam quality and brightness. This is not only true for edge-emitting lasers but also for vertical-cavity surface-emitting lasers (VCSEL) \cite{grabherr98,grabherr99,miller01,gronenborn11,orchard11,higuchi12}.
Properly heat-sinked devices with 100~$\mu$m diameter can reach an output power of nearly 0.4~W \cite{higuchi12}, 320~$\mu$m diameter nearly 1~W \cite{miller01}. This makes them attractive as inexpensive sources for medium brightness applications \cite{moench11}. From a fundamental point of view, the high Fresnel number of these devices makes them interesting to study optical self-organization in the plane transverse to the light propagation axes and it was shown that the instabilities can be funnelled into robust nonlinear excitations representing spatial solitons, first in a VCSEL with frequency-selective feedback \cite{tanguy07,tanguy08}, then in a VCSEL coupled to a saturable absorber \cite{genevet08}.

Laser cavity solitons (LCS) are a special class of dissipative solitons
in a system out of equilibrium where cavity losses are counterbalanced by the energy
input of the pumping \cite{akhmediev05,ackemann09a,barbay11}.
Since a laser can be driven by pump sources with
low temporal and spatial coherence or even completely incoherent sources such as
electrical pumping, cavity soliton lasers may be easier to operate than cavity soliton
schemes relying on coherent driving \cite{barland02}.
We are going to analyse in detail the formation of LCS in a VCSEL with frequency-selective feedback and provide a systematic investigation of the dependence of the properties of LCS on the ambient temperature and injection current. We will  demonstrate that the threshold condition is mainly determined by the detuning between the VCSEL resonance and the frequency-selective element, controlled by the device temperature. The latter is influenced by the ambient temperature and Joule heating. We will show that thresholds can differ by more than a factor of two changing the initial detuning when modifying the ambient temperature, without a significant change in the principal properties of the solitons except their power. At some point, however, high-order solitons - vortex solitons - can emerge as first structures above threshold. These vortex solitons were observed before in systems with saturable absorber \cite{genevet10a} and frequency-selective feedback \cite{jimenez13}.

LCS, as any self-localized structure, should have complete freedom to exist at any spatial location in a homogeneous system \cite{firth96,ackemann09a}. It is well known that in real semiconductor devices, spatial disorder - i.e.\ random spatial variations - of the  cavity resonance leads to a pinning of LCS at certain positions and also to a variation of threshold conditions \cite{tanguy08,genevet08,menesguen06,barbay08,pedaci08,pedaci08a,ackemann12}.
An understanding of these issues is very important if the bistable nature of LCS is to be exploited for optical processing applications, where they may limit the usefulness of cavity soliton lasers in spite of the high potential for parallelism created by the coexistence of many LCS within the broad aperture \cite{ackemann09a,barbay11}. We will provide numerical evidence that the disorder is responsible for the spontaneous switch-on of LCS as a control parameter (here current) is swept and that the variation of detuning parameter explains the spread of thresholds observed. We also establish that the shape of the LCS is not significantly influenced  by the trapping potential, i.e.\ that the observed LCS are close to the ones expected for the homogenous system, allowing their identification as solitons in spite of the underlying disorder.

\section{Experimental setup}

\begin{figure}[htb]
\centering
\includegraphics[width=\textwidth]{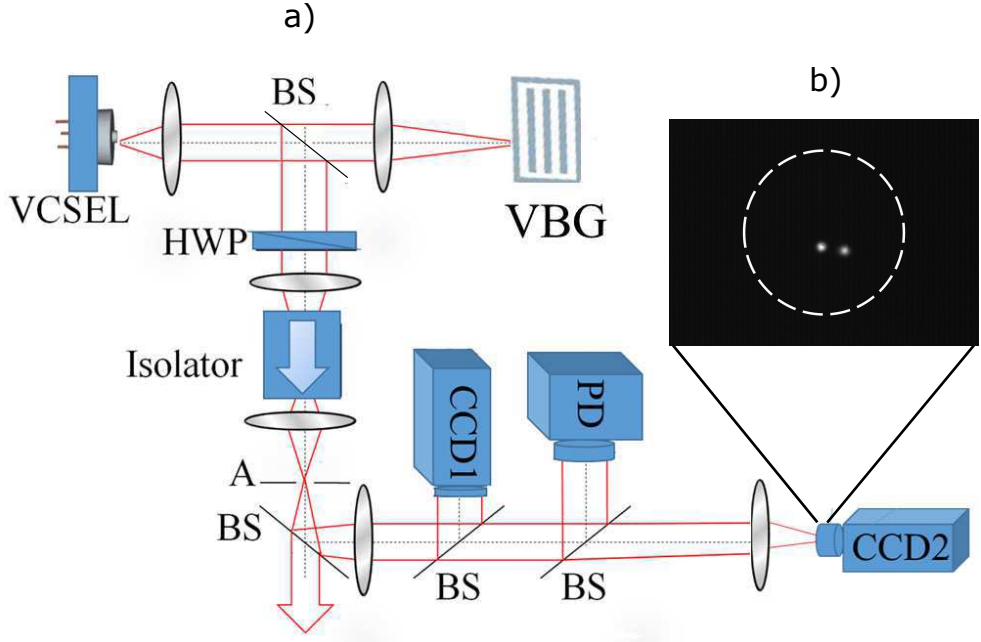}
\caption{(a) Experimental setup. VCSEL: Vertical-cavity surface-emitting laser, BS: Beam sampler, VBG: Volume Bragg grating, HWP: Half-wave plate, A: Aperture, PD: Photodiode, CCD2: charge coupled device (CCD) camera in near field image plane of VCSEL (magnification factor $50:1$), CCD1: CCD camera in far field image plane of VCSEL. (b) Typical near field intensity distribution of the VCSEL containing two laser cavity solitons, obtained at a low enough magnification factor to capture the whole VCSEL on the CCD sensor. The dashed white circle indicates the pumped aperture of the VCSEL with a diameter of  200~$\mu$m. }
\label{fig1}
\end{figure}

The setup is depicted schematically in Fig.~\ref{fig1}a. The VCSEL used  was
fabricated by Ulm Photonics and is similar to the ones described in
more detail in
\cite{grabherr98,tanguy08,grabherr99,radwell09}. It is a
large aperture device, allowing for the formation of many transverse cavity
modes of fairly high order, with a $200~{\rm \mu m}$ diameter
circular oxide aperture providing optical and current guiding.
The emission takes place through the n-doped Bragg reflector and the
transparent substrate (bottom-emitter, \cite{grabherr98,grabherr99}).
The laser has an emission wavelength around $975~{\rm nm}$ at room
temperature.  A volume Bragg grating (VBG) serves as frequency-selective element. It has a reflection peak of 95\% at
$\lambda_{\rm g}=978.1~{\rm nm}$ with a reflection bandwidth of $0.1~{\rm nm}$
full-width half-maximum (FWHM)  (see Fig.~3b of \cite{jimenez13}). The submount of the VCSEL is temperature controlled via a Peltier element and a feedback circuit in a range of $20-40^\circ$C.  In this  temperature
 range lasing is only possible with feedback from the VBG. The VCSEL is tuned in temperature so that the longitudinal cavity resonance approaches the reflection peak of the
VBG.  However,  the sensing thermistor is obviously not directly  in the active gain area of the VCSEL. Hence the Ohmic dissipation of the injection current still causes residual heating of the active zone. The cavity resonance of the VCSEL is measured to red-shift at a rate of 0.0035 nm/mA.

The VCSEL is collimated using an  aspheric lens with an effective focal length of  $f_1=8~{\rm mm}$, and  refocused down onto the VBG by a plano-convex lens with $f_2=50~{\rm mm}$ focal
length. This telescope setup gives a $6.25:1$ magnification factor onto the
VBG and is adjusted to be afocal, i.e.\ the external cavity is  self-imaging  after a round-trip. This   maintains the high Fresnel number of the VCSEL cavity and ensures local feedback, a prerequisite for self-localization.

A glass plate with an uncoated facet at the front and an anti-reflection coated facet at the back is used as a beam sampler to couple out the light from the cavity. The reflection is relying on Fresnel reflection and therefore is polarization dependent. The reflectivity is on the order of $10~{\rm
\%}$ for s-polarized light and $1~{\rm \%}$ for p-polarized light. The detection path is isolated from the cavity via an optical isolator serving also as a linear polarizer. We found that different solitons have slightly different polarization axes \cite{radwell09}, however, this is not important for the purposes of this report. Hence, we report data obtained for the vertical polarization component for all LCS. However, non-trivial polarization effects (polarization switching for fundamental solitons and inhomogeneous polarization for high-order solitons) have been observed recently at lower temperatures of the VCSEL submount, and are currently under investigation. After the isolator, there is an intermediate image plane with an adjustable circular aperture by which we can select a specific part of the VCSEL aperture for analysis.

 An amplified, slow photodetector (PD) is used to monitor the total emitted power. Charge-coupled device cameras (CCD)  allow us to observe and record the far field (CCD1) and near field (CCD2) intensity distributions. In order to be able to visualize the near field, another telescope is constructed. The total magnification used  is $50:1$, this gives us  a reasonable resolution of the soliton structure. As a result, only a part of the VCSEL aperture can be imaged onto CCD2 (cf.\ Fig.~\ref{fig1}c).

\section{Experimental observations and thermal properties of LCS}

\begin{figure}[htb]
\centering
\includegraphics[width=\textwidth]{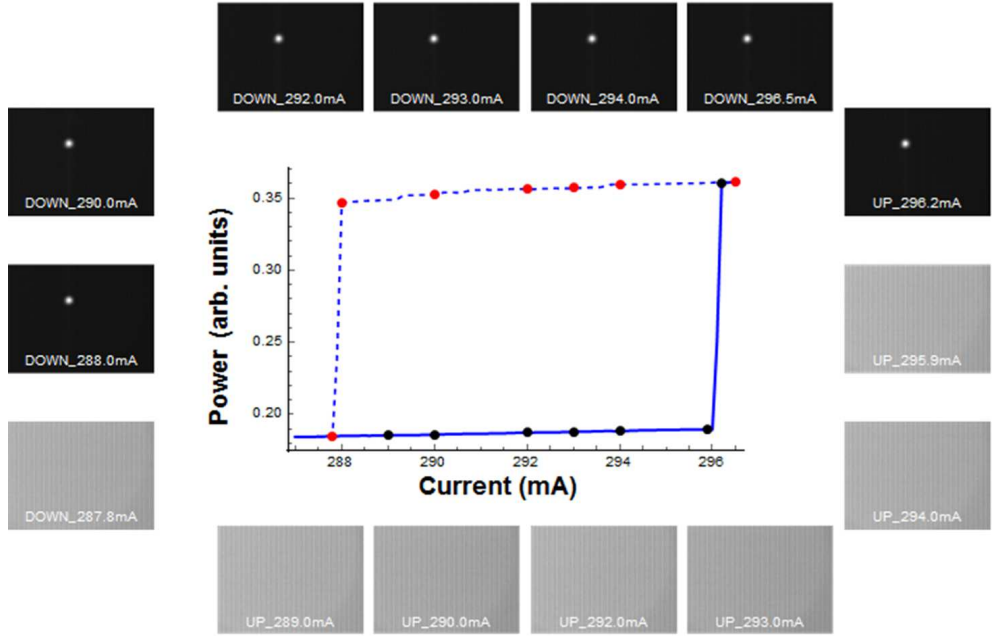}
\caption{Light-current characteristic and near field emission structures from the VCSEL.
The dots in the LI-curve denote the points were the images were taken increasing (black) and decreasing (red) the current. The current values are also in insets. The images are displayed in a linear grey scale with white
denoting high intensity. All images are obtained at the same gain and filter settings of
CCD2, but each image is scaled to obtain maximum contrast within the image. Hence the
images containing only spontaneous emission are dominated
by RF-interference.
Submount temperature: $40^\circ$C, displayed area: $171\times 127\,{\rm \mu m}$.}
\label{fig2}
\end{figure}

In Fig.~\ref{fig2} we illustrate the simplest scenario where the injection current into the VCSEL is increased. The light-current (LI) characteristic in the centre of Fig.~\ref{fig2} shows an abrupt increase of the emitted power from the background level given by spontaneous emission (lower row of images in of Fig.~\ref{fig2}) at a certain threshold  (about 296~mA). This increase of emission power does not take place over the whole laser aperture. Instead, a well defined spot of light (right upper inset in Fig.~\ref{fig2}) forms, which is much smaller than the device aperture (see Fig.~\ref{fig1}c for a direct comparison),  i.e.\ the switch-on from the off-state occurs to a localized state. This single-peaked spot structure is the fundamental LCS.

 If the current is decreased, an abrupt transition to the background state takes place, as typical for dissipative solitons. The fundamental LCS can coexist with the non-lasing state, as can be seen from the LI-curve in Fig.~\ref{fig2}, i.e.\ it is hysteric and bistable.  Within the coexistence range of the non-lasing state and the LCS, one can switch LCS on and off via external control beams \cite{tanguy07,tanguy08,radwell09}. This possibility to manipulate LCS makes them attractive for optical information processing, in principle \cite{firth96,ackemann09a,barbay11}.

\begin{figure}[htb]
\centering
\includegraphics[width=\textwidth]{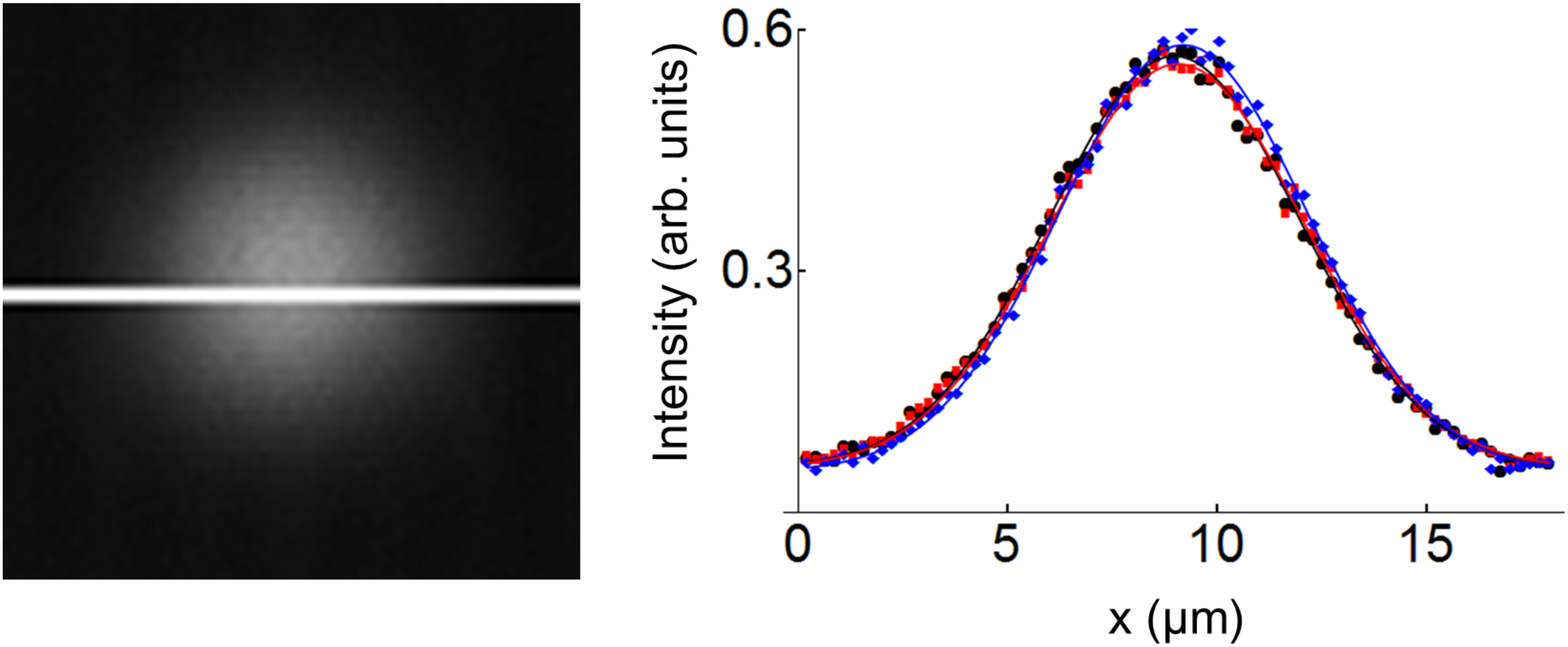}
\caption{(Color online) Left panel: Image of the fundamental LCS with the location of the cut for the intensity profiles presented in the right panel indicated by the horizontal white line. Injection current: 296.5~mA, VCSEL temperature for zero current: 40$^\circ$C. Right panel: The black circles  correspond to the intensity profile of the soliton after the switch-on (296.5 mA), the red squares to a profile in  the middle of the hysteresis loop (292 mA) and the blue rhombs to a profile just before switch-off (288 mA).}
\label{fig3}
\end{figure}

In Fig.~\ref{fig3} we provide an investigation on the shape of the fundamental LCS when the current is swept forward and backward. The figure displays the profiles of the  intensity  distribution from cuts  along an horizontal white line (left panel in Fig.~\ref{fig3}), taken at different values of the current scan. The black circles  correspond to the intensity profile of the soliton after the switch-on from the non-lasing state (296.5 mA). The red squares are obtained at a  current value in the middle (292 mA) of the hysteresis loop, the  blue rhombs just before the switch-off at  288 mA. The solid lines correspond to fitting curves of a Gaussian function giving a value of $6~{\rm \mu m}$ for the radius of the LCS at the $1/e^2$-points. The three profiles are rather similar indicating that the shape  of the LCS is quite independent of the injection current and whether it is bistable or not.

\begin{figure}[htb]
\centering
\includegraphics[width=\textwidth]{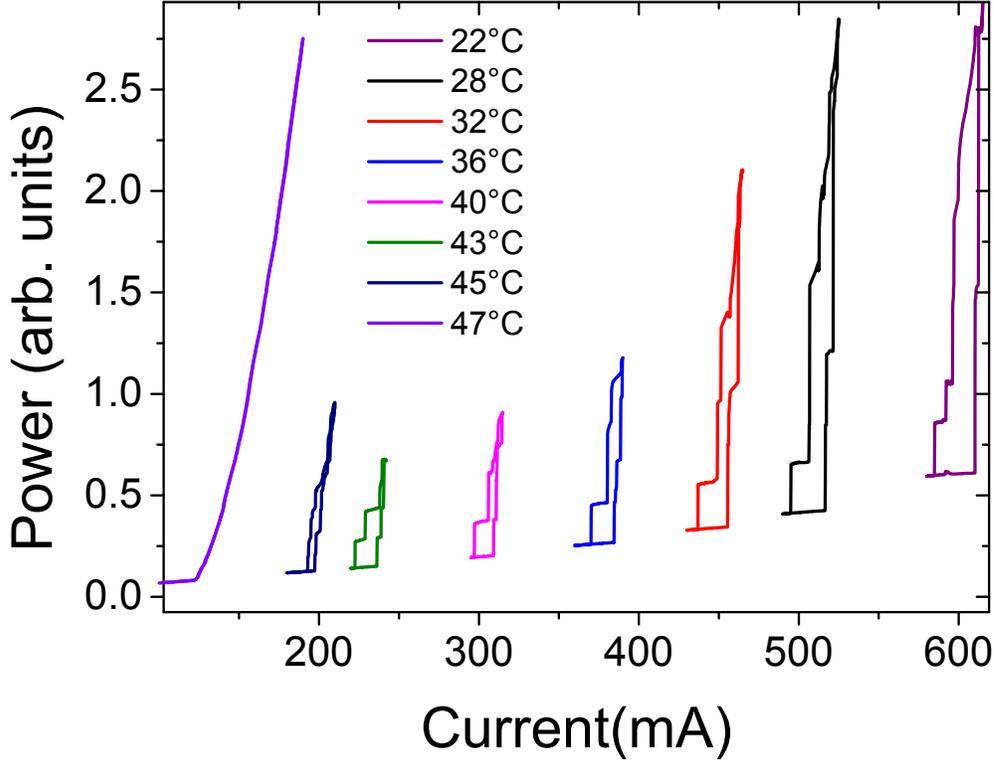}
\caption{(Color online) a) LI-curves obtained for different submount temperatures (in legend). The submount temperature decreases from left to right from $47^\circ$C to $22^\circ$C via the steps indicated in the legend. The LI-curves in the centre of the figure correspond to situations where the fundamental LCS is the first structure emerging, whereas the last one on the left shows a continuous switch-on to a non-solitonic extended state ($47^\circ$C) and and the last one on the right displays the switch-on to a vortex soliton with a higher total power than the fundamental LCS ($22^\circ$C).}
\label{fig4}
\end{figure}

Fig.~\ref{fig4} presents an analysis of the LI-curves at different submount temperatures. Abrupt switching and hysteresis occur over a large temperature interval. However, at high temperatures ($>47^\circ$C) the LI-curves  are actually continuous, i.e.\ do not show bistability any more, but the switch-on behavior of a conventional laser. Within the bistable range,  main observations  are that  the switch-on and the switch-off current thresholds are increased considerably and the width of the hysteresis loop widens, when the  submount temperature is decreased. Furthermore, the LCS power at switch-off and switch-on increases with switching current (lower temperature). At low temperature (22$^\circ$C), the initial jump is to a much higher amplitude as larger structures than the fundamental soliton switch on at threshold (see below, Fig.~\ref{fig8}).

\begin{figure}[htb]
\centering
\includegraphics[width=\textwidth]{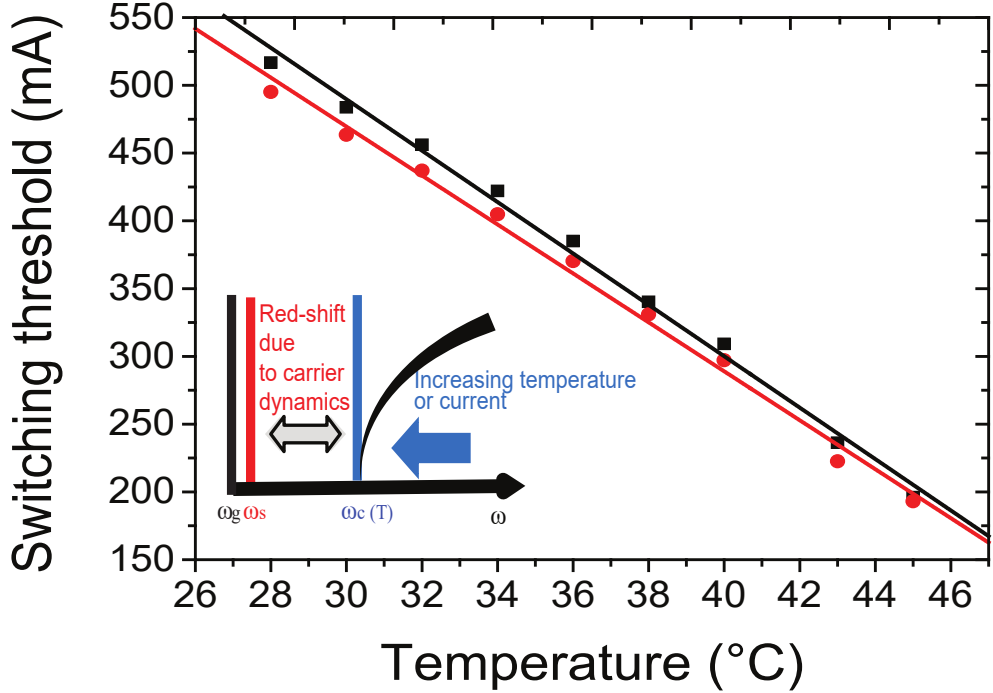}
\caption{(Color online) Switch-on (black squares) and switch-off thresholds (red circles) in
dependence on submount temperature. The straight lines are linear fits to the data. The inset shows an illustration of detuning conditions. The cavity resonance ${\rm\omega_c}$(T)  red-shifts towards the VBG resonance ${\rm\omega_g}$ as the current or submount temperature is increased. Then at a some temperature or current, abrupt closure of the gap takes place due to the intrinsic carrier related nonlinearity of the semiconductor. The dispersion relation of the VCSEL is broadened by  disorder. ${\rm\omega_s}$ is the resulting LCS frequency.}
\label{fig5}
\end{figure}

In order to have  a quantitative description of the switch-on and switch-off points we  plot
the current thresholds against the submount temperature (Fig.~\ref{fig5}). The thresholds
decrease approximately linearly when the temperature increases. Fitting yields a slope of $(-19.0\pm0.5)~{\rm mA/K}$ for the switch-on thresholds and $(-18.1\pm0.6)~{\rm mA/K}$ for the switch-off thresholds. The
small difference is due to the widening of the hysteresis loops as discussed earlier.
From previous measurements without feedback, we know how much the wavelength of
the cavity resonance shifts with submount temperature ($d{\rm\lambda_c}/dT~{\rm\approx}~0.066~{\rm nm/K}$)
and with current ($d{\rm\lambda_c}/dI~{\rm\approx}~0.0035~{\rm nm/mA}$). Henceforth, one estimates a heating rate of the device by Ohmic losses as ($dT/dI~{\rm\approx}~0.053~{\rm K/mA}$). Finding the inverse of this yields $dI/dT~{\rm\approx}~18.9~{\rm mA/K}$. This rate is very close to the one measured for the change in switch-on current and still very close to the rate of change of the switch-off current. This result can be interpreted as the active zone of the VCSEL  switching always at approximately the same temperature, independently of the gain. This temperature can be estimated from the measured thresholds and the known shifts to be about 56$^\circ$C. It is somewhat decreasing (by about 0.5$^\circ$) for low temperatures (high threshold currents), which is reflected by the slight deviation of the characteristic in Fig.~\ref{fig5} from a straight line in this region. Before drawing conclusions from this deviation, it should be cautioned that the temperature and the corresponding wavelength shift with current do not need to be linear over large current changes but can contain a quadratic component as Joule heating is proportional to the square of the current. This can also account for the slight deviation and switching might still occur at constant active area temperature. Switch-on and switch-off thresholds are separated by about 1$^\circ$C.

This kind behaviour can be illustrated by the scheme in the inset  in Fig.~\ref{fig5}. At high temperatures, the resonance of the VBG is within the mode spectrum of the VCSEL and the result is a continuous switch-on of a spatial mode with the wavelength selected by the VBG. Also switch-on with feedback from a plane mirror replacing the VBG occurs at about 180~mA \cite{noblet12}. However, at low temperature  at the beginning (zero current) the longitudinal resonance of the VCSEL is at higher frequency than the reflection peak of the VBG and there is no feedback. Increasing the device temperature (either by increasing the submount temperature or increasing the Joule heating via the current) red-shifts the VCSEL resonance  and it moves closer to the VBG resonance. If the gap is small enough (about 0.06 nm or 20 GHz inferred from the difference in switching temperatures of about 1 degree), nonlinear effects in presence of fluctuations can lead to spontaneous switch-on.  A small fluctuation increasing emission intensity leads to a decrease in carrier number due to stimulated emission. In semiconductor lasers the refractive index is strongly dependent on carrier density. Their refractive index increases with decreasing carrier number. For a simplest description often a linear relationship is assumed with the proportionality constant given by Henry's ${\rm\alpha}$-factor \cite{henry82}. As the cavity resonance of a Fabry-Perot laser is given by $\omega_c =c/(2Ln)$, an increase in emission will lead to an increase in refractive index thus decreasing the cavity resonance. Hence the VCSEL resonance moves closer to the VBG frequency, the reflectivity of the external cavity increases, the losses of the system become smaller and due to positive feedback intensity will increase further and abrupt switching takes place. This mechanism corresponds to dispersive optical bistability \cite{lugiato84}.

\begin{figure}[htb]
\centering
\includegraphics[width=\textwidth]{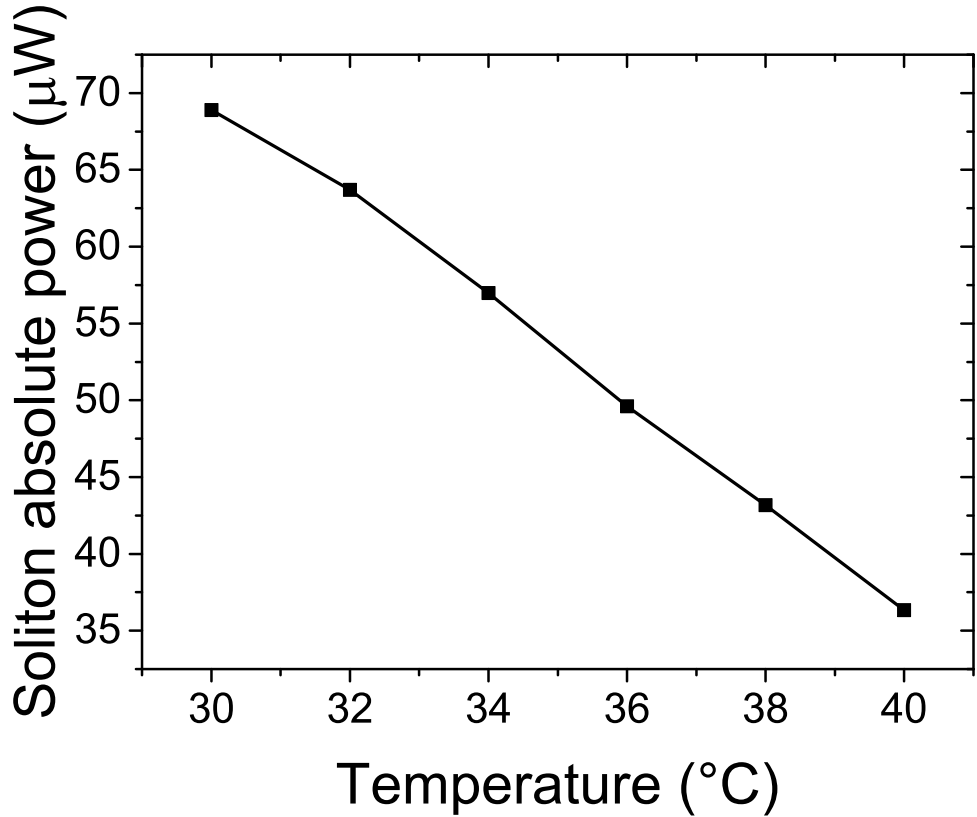}
\caption{Dependence of absolute power of LCS on submount temperature in the range of $30-40^\circ$C. The lines are guiding the eye. These measurements were taken on a different day than the previous figures, with slight changes in the alignment of the VBG.}
\label{fig6}
\end{figure}

 From Fig.~\ref{fig4} one could analyze power and amplitude of the soliton, but we prefer to present a direct measurement in order to discuss absolute values.
 A powermeter measuring the total output power of the VCSEL is placed after the BS. We measure the power just above the LCS switch-off threshold on up and down scans of the injection current. The difference constitutes the soliton power (Fig.~\ref{fig6}).  The absolute power is some tens of $\mu$Ws. From the reflectivity of the BS and the output coupler (estimated to be 0.9977), one can estimate the power within the VCSEL to be about 250 mW for 50~$\mu$W of outcoupled power, enough for substantial nonlinear effects. The LCS power increases by about a factor of two over a range of ten degrees. This can be qualitatively explained as the gain is higher at lower ambient temperatures due to the higher threshold current. (Note that the temperature of the active zone is approximately constant as argued earlier.) It is also easily conceivable that a high gain enables to cross larger gaps, which explains qualitatively the observed tendencies  to larger hysteresis loops as well as possibly the deviation of the switching characteristics  in Fig.~\ref{fig5} from a straight line. In conclusion, switching is dominantly reigned by detuning changes due to thermal and carrier dynamics but details are influenced by the gain.

\begin{figure}[htb]
\centering
\includegraphics[width=\textwidth]{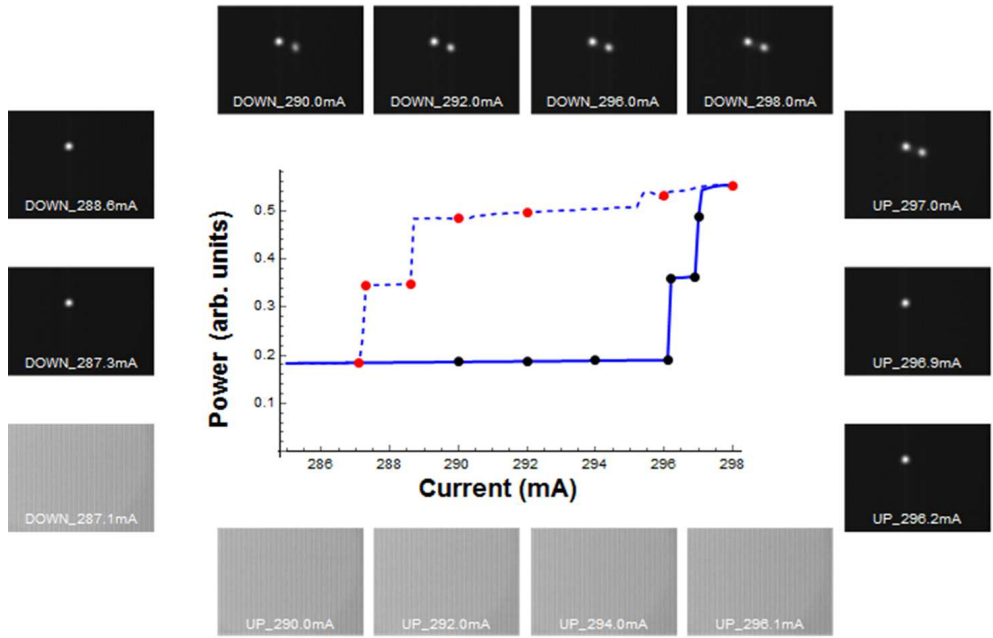}
\caption{Light-current characteristic and near field emission structures of the VCSEL. Same explanation and parameters as in  Fig.~\ref{fig2} but  the injection current is increased a little further, showing switch-on of a second LCS  at a different location of the VCSEL. The small kinks in the LI-curve are due to longitudinal mode-hops. Submount temperature: $40^\circ$C.}
\label{fig7}
\end{figure}

Fig.~\ref{fig7} gives an example of what is typically happening at  high temperatures, if the current is increased beyond the hysteresis loop of the first LCS in Fig.~\ref{fig2}. A further LCS appears at a different location, again in an abrupt fashion. When reversing the direction of the scan, bistability between the states with a  single and two solitons is obtained. Decreasing the current further, there is bistability between the non-lasing background and two solitons. For even lower current, the second LCS switches off abruptly. Further decreasing the current, the  remaining scenario follows the one of Fig.~\ref{fig2}.  The difference in switching currents is about 1~mA, corresponding to about 0.0035 nm or 1 GHz.

From the discussion of Fig.~\ref{fig7}, the dispersion in LCS thresholds suggests a dispersion of detuning conditions and hence a spatial variation of the VCSEL resonance. This is due to uncontrolled fluctuations in the epitaxial growth process, which cause a variation of resonance conditions across the pumped aperture of the device. These growth fluctuations in semiconductor microcavities are well known to influence
linear effects like coherent backscattering \cite{gurioli05}, but in particular also
nonlinear phenomena relying on resonant field enhancement \cite{oudar92,sanvitto06} and limit the coherence of VCSEL-arrays \cite{pier97}. An additional consequence for LCS is that they are pinned: In a homogeneous system one expects that a self-localized state like a soliton  could exist at any location. In  the experiment, it is observed that the LCS are appear preferentially at certain locations. These locations depend on device and can be changed slightly, but only slightly, by alignment of the VBG \cite{radwell09}.  Translation is a Goldstone mode of a soliton. It will couple to all spatially inhomogeneous perturbations and the LCS will move until it reaches a local extremum of the perturbation, where the gradient vanishes \cite{rosanov91,firth96,taranenko97,schaepers01}. This leads first to a pinning of the solitons at certain positions generally referred as either traps or defects. This was investigated in detail in coherently driven semiconductor microcavities \cite{pedaci08,pedaci08a} and by one of the authors in the VCSEL with FSF  \cite{ackemann12}. These investigations further support the approach taking switching threshold as an indication of detuning proposed in  \cite{ackemann12} as well as identify potential limitations, if the current range investigated is large. From an applicative point of view, this disorder  constitutes a major obstacle to  exploiting  the potentially massive parallelism offered for information processing by the coexistence of many CLS in a single broad-area cavity soliton laser, as it turns out that monolayer fluctuations are enough to affect CLS behaviour significantly \cite{oudar92,ackemann12}.

\begin{figure}[htb]
\centering
\includegraphics[width=\textwidth]{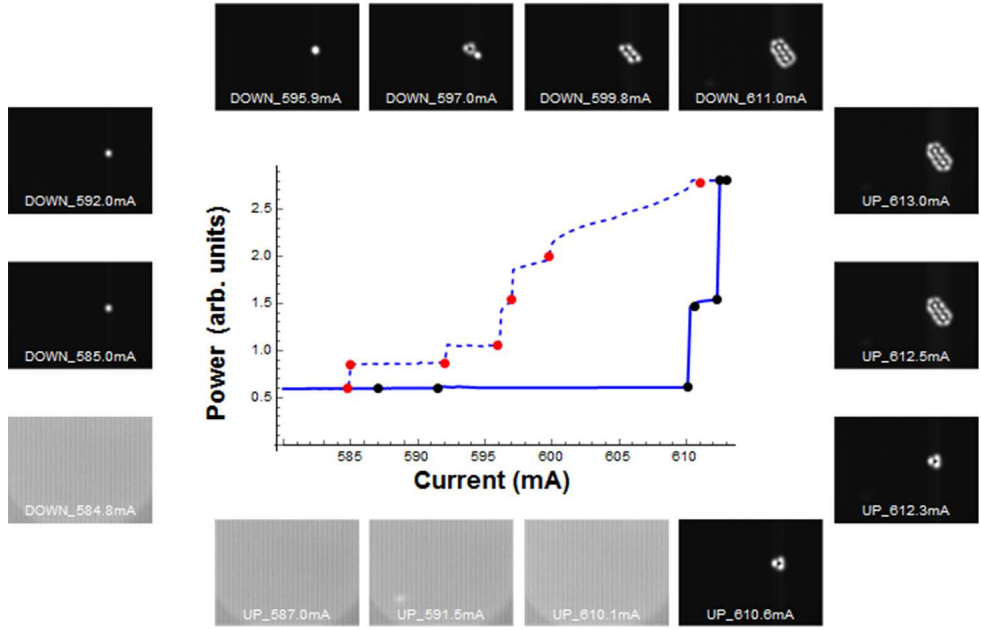}
\caption{Light-current characteristic and near field emission structures of the VCSEL at low submount temperatures. Switch-on occurs to a high-order soliton. Submount temperature: $22^\circ$C.}
\label{fig8}
\end{figure}

As the last observation, a novel result related with the first structure formed spontaneously from the nonlasing state is reported in Fig.~\ref{fig8}. It shows the LI-curve of the device and typical images for essentially the same situation as Fig.~\ref{fig2} and Fig.~\ref{fig7} but at a much lower temperature of $22^\circ$C. The scenarios at 40$^\circ$ and 22$^\circ$C are very similar regarding the existence of solitons, abrupt switching and bistability, but details differ significantly. Now the first soliton state switching-on from the non-lasing state is  a ring structure with some amplitude modulation along the azimuthal angle. This structure was identified as a vortex soliton in \cite{jimenez13} by interference measurement. The deviation of the intensity structure from the ring is currently under investigation and hints to the possible interpretation as azimuthon, i.e.\ a vortex with an azimuthal modulation of intensity with a symmetry different from the order of the phase structure of the vortex. Corresponding structures were predicted for conservative \cite{desyatnikov05a} and dissipative systems \cite{fedorov03,soto-crespo09} and related structures were observed in quasi-conservative systems \cite{minovich09}.

At higher current, the vortex soliton gives way to more complex structures, which however remain localized and are different from the extended wavy high order modes of the VCSEL itself \cite{noblet12}. The details of the intermediate states of these transition between vortices to larger states and then back to the vortex depend sensitively on temperature, minute alignment of the VBG and how far the initial up-scan of current was performed, but the common feature for the lower order states is that the vortex soliton can switch on directly from the off-state, but will switch down to the fundamental LCS as an intermediate state before the final switch-off to the non-lasing state. For even lower temperatures a structure with an even larger spatial extension displaying a central peak and one ring around it can appear at threshold. We did not perform systematic investigations on this as the high currents are stressing the laser. It is worth to mention that at temperatures close to $20^\circ$C  the spontaneous formation of non-trivial polarization states of the fundamental soliton and high-order soliton have been also observed and are under investigation. For example, the jump at 592~mA in the upper branch of the LI-curve in Fig.~\ref{fig8} is due to a polarization switching of the fundamental LCS.

\section{Theoretical analysis}

For a theoretical analysis of the experimental observations, we are using a class-B VCSEL model that includes carrier dynamics and frequency-selective-feedback \cite{paulau09}:
\begin{eqnarray}
\label{model}
\partial_\tau E &=&  -(1+i\theta) E + (1-i \alpha) NE + F + i \nabla^2 E \nonumber \\
\partial_\tau N &=& \gamma \left[ \mu - N \left( 1 + |E|^2 \right) \right] \\
\partial_\tau F &=& - \lambda F + \sigma \lambda E \,\, ,\nonumber
\end{eqnarray}
where $E$, $N$ and $F$ are the intracavity electric field, the carrier density and the feedback field, respectively. $\theta$ is proportional to the difference between the cavity and grating frequencies, $\alpha$ the linewidth enhancement factor, $\gamma$ the ratio between the carrier and cavity decay rates, $\mu$ the injection current (gain), $\lambda$ the filter bandwidth, $\sigma$ the feedback strength and the time $\tau$ has been normalized via the photon lifetime in the cavity. In the simulations reported below we have kept the parameter values $\alpha=5$, $\gamma=0.01$, $\lambda=0.0271$ and $\sigma=0.6$ fixed. The control parameters are the detuning $\theta$ and the gain $\mu$. The model equations (\ref{model}) are equivalent to those provided in Ref. \cite{paulau09} apart from $E$ being the complex conjugate field, $\theta=\omega_s/\kappa-\alpha$ and the neglect of the delay time in the external cavity. We confine to a scalar model as polarization aspects are not important for the understanding of the current observations.

We have performed numerical simulations in one transverse dimension and described defects in the VCSEL background as local spatial modifications of the cavity detuning $\theta$. We have used two kinds of 1D profiles for the traps that simulate defects: rectangular and parabolic, with similar results. When ramping up the current, the simulations account for increasing gain and also, more importantly, for the sweeping of the VCSEL cavity resonance with respect to the feedback frequency. This has been simulated via a linear relation of the cavity detuning $\theta$ and the current $\mu$:
\begin{equation}
\label{tvsmu}
\theta = - 5.23 \mu + \theta_0
\end{equation}
as the cavity frequency decreases with increasing temperature. The value of the slope has been obtained to match experimental results. For the interpretation of the steady-state structures, a static temperature background is sufficient. Radiative cooling might provide a quantitative change to the detuning conditions, but can be thought of being incorporated into the disorder landscape, as long as details of the transients and the hysteresis width are not sought for.

\begin{figure}[htb]
\centering
\includegraphics[width=\textwidth]{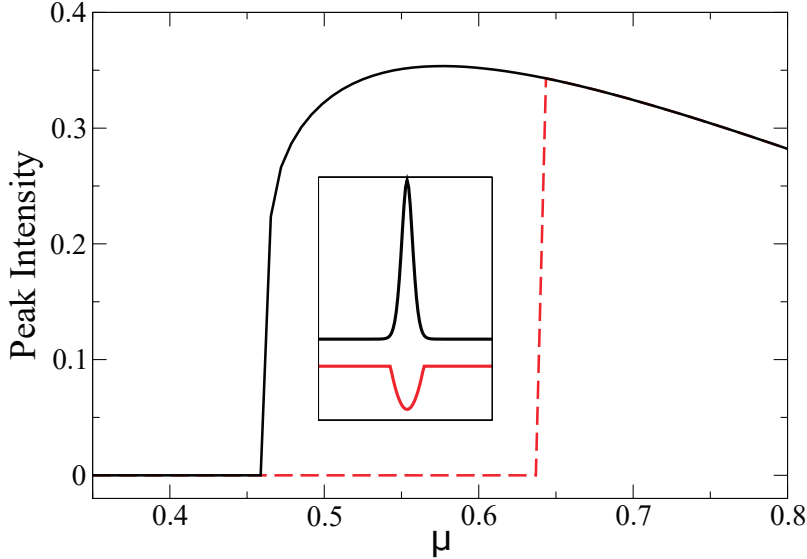}
\caption{Output power for a forward (dashed red line) and a backward (black solid line) scan of the gain parameter $\mu$ for the model (\ref{model}). The inset shows a sketch of the amplitude of the LCS for $\mu=0.55$ during the backward scan (black line), and the shape of the detuning trap (red line). The trap depth is $\Delta \theta = 1.6$. Other parameters are $\theta_0=1.77$.}
\label{fig01T}
\end{figure}

Fig.~\ref{fig01T} shows that spatial variations of the cavity tuning
can lead to spontaneous generation of LCS without the use of external writing beams as we demonstrated experimentally above. Upon increasing the current $\mu$, the background becomes unstable at around $\theta \approx -2.8$ and without the trap provided by the defect an extended periodic spatial structure would appear at this value of the effective detuning. If a sufficiently large area of the trap centre is around or beyond this detuning threshold, the instability is still triggered but the defect traps one peak (or few peaks) of the pattern state (see dashed red line in Fig.~\ref{fig01T}). Upon inverting the direction of the scan, the pattern peak evolves into a LCS (see the black line in Fig.~\emph{}\ref{fig01T}) that is bistable with the off-solution as observed in the experiment too.

\begin{figure}[!htb]
\centering
\includegraphics[width=15cm]{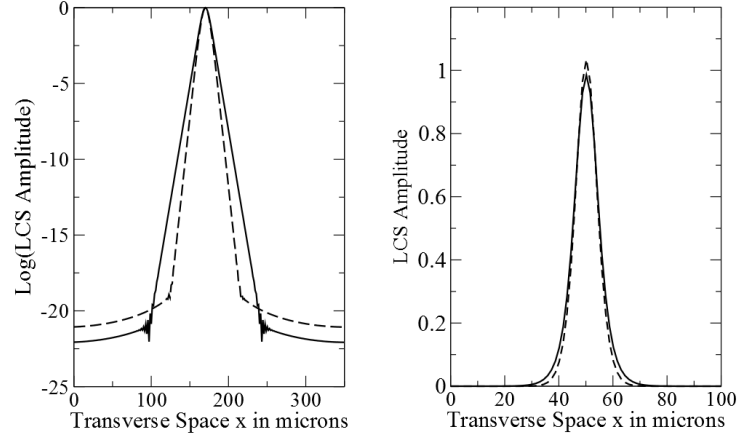}
\caption{Amplitude profile of a LCS on a flat background (solid line) and in a parabolic trap (dashed line) in a normal (left panel) and logarithmic scale (right panel). The maximum amplitude of the LCS without trap has been normalized to one. Parameters are the same as those in Fig.~\ref{fig01T} with $\mu=0.55$.}
\label{fig02T}
\end{figure}

In order to check that the localized structure in the trap is a LCS, we show  the amplitude profile of a LCS on a flat background (solid line) and that of the LCS in the parabolic trap (dashed line) in Fig. \ref{fig02T}. The two LCS are barely distinguishable from each other in linear scale, while on a logarithmic scale they show differences in the exponential tails, with the trapped LCS displaying a tighter confinement than the LCS on a flat background.

\begin{figure}[htb]
\centering
\includegraphics[width=\textwidth]{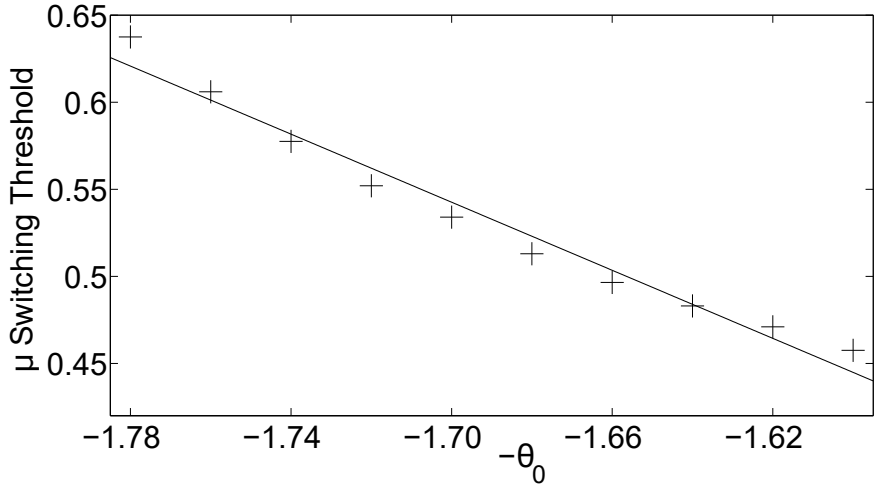}
\caption{Current threshold $\mu$ leading to the formation of a LCS in a parabolic trap versus the intercept $-\theta_0$ of Eq.~(\ref{tvsmu}). The crosses correspond to results from the numerical simulations while the solid line is a linear fit to the data. We use $-\theta_0$ in this plot for a comparison of the theoretical results with those of Fig. \ref{fig5}.}
\label{fig03T}
\end{figure}

By modifying the intercept $\theta_0$ of the linear dependence between gain and detuning given by Eq. (\ref{tvsmu}) we have simulated the temperature dependence of the threshold values of the gain $\mu$ for the generation of single peak LCS in a parabolic trap (see Fig.~\ref{fig03T}). A decrease in the magnitude of the initial detuning $\theta_0$, mimicking the VCSEL resonance shift due to submount temperature, reduces the injection current necessary for the appearance of the LCS.

\begin{figure}[htb]
\centering
\includegraphics[width=\textwidth]{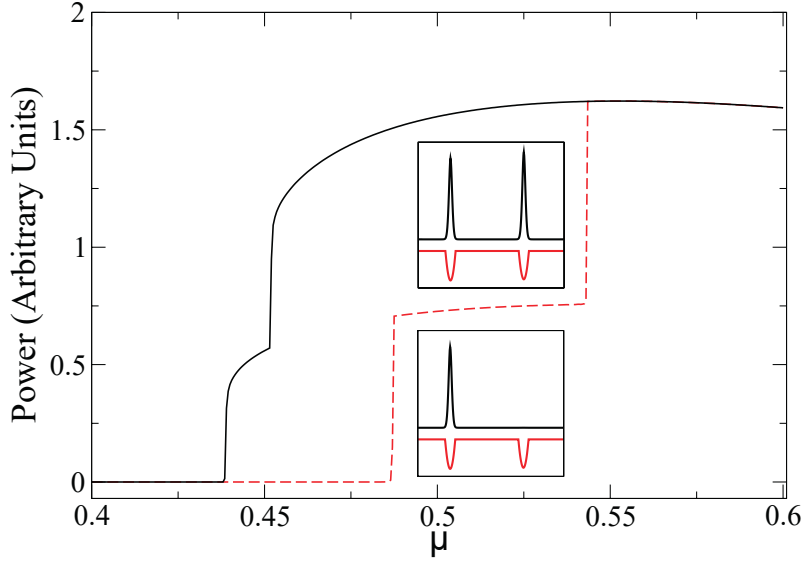}
\caption{Output power for a forward (dashed red line) and a backward (black solid line) scan of the gain parameter $\mu$ for a situation with two traps of slightly different depth ($\Delta \theta = 1.82$ for the trap on the left and $\Delta \theta = 1.75$ for the one on the right). The insets show a sketch of the field amplitude (black line) and trap profiles (red lines) during the forward scan (lower inset panel), and during the backward scan (upper inset panel), both for $\mu=0.5$. Other parameters are $\theta_0=1.8$.}
\label{fig04T}
\end{figure}

In analogy with the experiments presented in Fig.~\ref{fig7}, we have also simulated the spontaneous appearance of single peak LCS in two traps at a distance of few LCS sizes. Fig.~\ref{fig04T} shows a typical result of the forward and backward scans when changing the gain parameter $\mu$. Since the two parabolic traps have a slight difference in their depths, the pattern peak forms first in the left trap and then in the right trap during the forward scan (see the dashed red line in Fig. \ref{fig04T}). During the backward scan both peaks transmogrify into spatially separate single-peak LCS, when bistable operation with the off state occurs (for $\mu < 0.485$ in Fig.~\ref{fig04T}). Since both LCS are on, the total power is higher. The LCS on the right side, the second one to switch on, is the first one to switch off during the backward scan in qualitative agreement again with what has been observed experimentally in Fig.~\ref{fig7}. By comparing Fig.~\ref{fig04T} with Fig.~\ref{fig01T}, it is also apparent that the power of the LCS decreases with decreasing detuning, as experimentally observed.

\begin{figure}[htb]
\centering
\includegraphics[width=\textwidth]{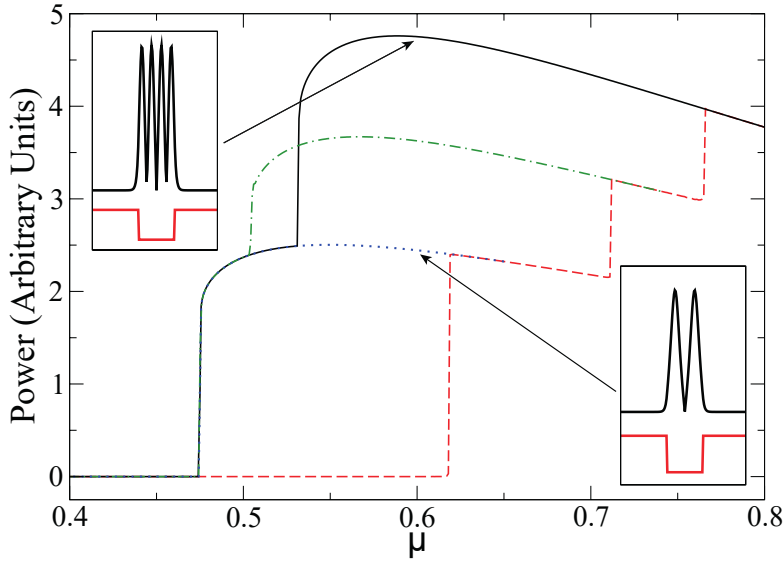}
\caption{Output power for a forward (dashed red line) and a backward (black solid line) scan of the gain parameter $\mu$ for the model (\ref{model}). The dash-dotted (green) line shows the backward scan started at $\mu=0.73$ instead of $\mu=0.8$ leading to the triple-peak LCS. The dotted (blue) line shows the backward scan started at $\mu=0.65$ leading to the triple peak LCS. The insets show a sketch of the amplitudes of the LCS for $\mu=0.6$ during the backward scans, and the shape of the detuning trap (red line). The trap depth is $\Delta \theta = 1.5$. Other parameters are $\theta_0=1.9$.}
\label{fig05T}
\end{figure}

For larger negative detunings (lower temperatures of the environment), we observe the appearance of multi-peak structures directly from the instability of the background when increasing the gain parameter in the forward scan. In Fig.~\ref{fig05T}, a double-peak structure is the first one to appear in a wide, flat trap followed progressively by structures with more and more peaks. Once the scan direction is reversed, these multi-peaked structures evolve into multi-peaked LCS.
In Fig.~\ref{fig05T} the double-peak structure is followed by the appearance of a triple-peak and then a four-peak structure during the forward scan of the gain parameter $\mu$. During the backward scan, the four-peak structure converts into a four-peak LCS at around $\mu$=0.62 where bistability with the background is observed. Before being re-absorbed by the background at the end of the backward scan of the gain parameter, the four-peak LCS decays into a double-peak LCS first and then disappears (see the solid black line in Fig.~\ref{fig05T}). Different sequences of multi-peaked LCS during the backward scans, such as from four to three to two to zero peaks or from four to two to one to zero peaks, can be observed when changing the shape and symmetry of the trap profile. For completeness, we show in Fig.~\ref{fig05T} the backward traces of the triple and double-peak structure (and corresponding LCS) if the reversing of the scan is started before the formation of the four-peak structure. The LCS formation and and behaviour displayed in Fig. \ref{fig05T} compare well, qualitatively, with Fig.~\ref{fig8} of the experimental results. We stress that the same trap supports only one LCS at threshold at lower absolute values of $\theta_0$, reproducing qualitatively the difference in observation between Figs.~\ref{fig2} and \ref{fig8}. In variance with the experiment, the two-peak LCS state switches down as a whole, whereas the typical observation in the experiment is that the switch-off occurs via the fundamental LCS (Fig.~\ref{fig8}). This difference might be due to the different dimensionality of the numerical simulations. Work in this direction is currently under way, assessing also the potential relevance of an imperfect rotational symmetry of the traps for this transition.

We mention that the hysteretic transition between different number of pulses within a round-trip time are also known for temporal dissipative solitons, e.g.\ in fibre lasers, and is usually linked to the total energy available in the cavity \cite{tang99,gutty01,tang05,liu10,grelu12,marconi14}. Whereas the transition between different soliton states in different traps is related to the disorder in spatial systems, the observation of states of different complexity and number of constituents in the same trap as a function of pump bears some similarities to the temporal case. For example, multi-soliton complexes will use the gain more effectively.

\section{Conclusion}
We have experimentally and theoretically provided a detailed characterization of the influence of the device temperature of the switch-on, switch-off and nature of the localized structures that appear in  traps in VCSELs with frequency-selective feedback. The investigations clear show that the decisive  effect of the device temperature is to control the detuning between VCSEL and VBG resonance, which determines the threshold. The device temperature is in turn controlled by the submount temperature and the current induced Joule heating. To first approximation, the latter is the decisive driver to induce LCS switch-on by changing the current, the increase in gain due to the higher carrier number mainly controls soliton amplitude, not threshold. This is well reproduced in a numerical model taking into account the double effect of current on gain and detuning. We also provided first numerical characterization of the influence of detuning disorder on thresholds and the possibility of a spontaneous switch-on of a localized spot in a trap without an external perturbation. Although this spot is only a part of an extended structure localized by the trap at the spontaneous switch-on, it transforms into an essentially identical state representing the LCS, if the driving is decreased again into the bistable range. This provides strong support for the claim that the experimentally observed structures in the system with disorder are a manifestation of LCS, within the limits on choice of position and dispersion of threshold imposed by the disorder.

For large initial detuning (low temperatures in the experiment), high-order solitons (vortex solitons in the 2D experiment, closely packed multi-humped structures in the 1D simulations) can appear as the first structure emerging at threshold.  The first investigation on vortex solitons reported these structures at 39$^\circ$C \cite{jimenez13} and either the fundamental LCS or the vortex were selected depending on minute alignment changes of the VBG, interpreted as inducing small relative shifts of feedback phase and detuning conditions. The present investigations clarify that the fundamental LCS is the preferred structure at small initial detuning and low thresholds, whereas the vortex or even larger structures are the preferred ones at large initial detuning and large current. A tentative interpretation is that at higher gain larger structures can be sustained in the same trap. Theoretically, only very limited guidance for the selection is available. In \cite{jimenez13} the situation was analyzed using a class A Ginzburg-Landau-like model. The bifurcation analysis indicates that the fundamental LCS and the vortex soliton are nearly frequency degenerate but that the fundamental solitons start to emerge slightly sooner from a saddle-node bifurcation, if the detuning is decreased at constant current, i.e.\ constant gain. Work is presently done to extend the simulations with the coupling between detuning and gain to two transverse dimensions. Obviously, a thorough understanding of the interplay of disorder (including potentially the absence of rotational symmetry in the traps), gradients of feedback phases due to minute misalignment of the VBG and nonlinearity is very important for the operation of real devices, but very challenging to analyse experimentally as well as theoretically.

\section*{Acknowledgment}J.G. gratefully acknowledge support from
CONACYT.  We are grateful to Roland J\"ager (Ulm Photonics) for supplying the devices, and to Alison Yao for useful discussions and help in the preparation of the figures, and to W. J. Firth for numerous fruitful input and discussion on soliton frequency shifts and the spontaneous appearance of solitons.

\section*{References}


\providecommand{\newblock}{}

\end{document}